\documentclass[a4paper]{jpconf}
\usepackage{amsmath}
\usepackage{amssymb}
\usepackage{graphicx}
\usepackage[mathscr,scaled=1.15]{urwchancal}
\DeclareFontFamily{OT1}{pzc}{}
\DeclareFontShape{OT1}{pzc}{m}{it}%
{<-> s * [1.15] pzcmi7t}{}
\DeclareMathAlphabet{\mathpzc}{OT1}{pzc}{m}{it}

\begin{document}
\title{Hadron Physics and QCD: Just the Basic Facts.}

\author{Crag D.\ Roberts}

\address{Physics Division, Argonne National Laboratory, Argonne, Illinois 60439, USA}

\ead{cdroberts@anl.gov}

\begin{abstract}
With discovery of the Higgs boson, the Standard Model of Particle Physics became complete.  Its formulation is a remarkable story; and the process of verification is continuing, with the most important chapter being the least well understood.  Quantum Chromodynamics (QCD) is that part of the Standard Model which is supposed to describe all of nuclear physics and yet, almost fifty years after the discovery of quarks, we are only just beginning to understand how QCD moulds the basic bricks for nuclei: pious, neutrons, protons.  QCD is characterized by two emergent phenomena: confinement and dynamical chiral symmetry breaking (DCSB), whose implications are extraordinary.  This contribution describes how DCSB, not the Higgs boson, generates more than 98\% of the visible mass in the Universe, explains why confinement guarantees that condensates, those quantities that were commonly viewed as constant mass-scales that fill all spacetime, are instead wholly contained within hadrons, and elucidates a range of observable consequences of confinement and DCSB whose measurement is the focus of a vast international experimental programme.
\end{abstract}

\section{Introduction}
A context for this report is provided by the experimental programmes underway and planned at the Thomas Jefferson National Accelerator Facility (JLab), Newport News, Virginia.  Funds for the development of research plans and designs for the Continuous Electron Beam Accelerator Facility (CEBAF) at JLab were initially provided in 1984, and construction began in 1987.  Seven years later, in 1994, the facility achieved its design capability, delivering 4\,GeV electron beams on targets in the three associated experimental halls.  The goal was to ``write the book'' about the strongest known force in Nature -- the force that holds nuclei together -- and determine how that force can be explained in terms of the gluons and quarks of QCD.

An aim of the original JLab programme was verification of the following prediction: at energy-scales greater than some minimum value, $Q_0$, whose value was not determined by the proof, cross-sections and form factors involving hadrons should behave as follows: \cite{Brodsky:1973kr,Brodsky:1974vy,Farrar:1979aw,Efremov:1979qk,Lepage:1979zb,Lepage:1980fj}
\begin{equation}
\label{QCDtest}
{\mathpzc A}(\sigma^2) \stackrel{\sigma\gg Q_0}{\propto}
\left[\frac{Q_0^2}{\sigma^2}\right]^{{N}}
\ln\left[\frac{Q_0^2}{\sigma^2}\right]^{\gamma_{\mathpzc A}}
\end{equation}
where the integer $N=n_{\rm quarks} -1 + \Delta\lambda$, with $n_{\rm quarks}$ counting the lowest possible number of quarks and/or antiquark in the target and $\Delta\lambda=0,1$, depending upon whether a spin-flip occurs in the scattering process.  The power-law term gives rise to parton-model scaling and can be explained on dimensional grounds.  The distinctive signature of QCD lies in the logarithmic factor, the exponent on which, ${\gamma_{\mathpzc A}}$, can be computed and whose appearance signals scaling violation, which is well documented in deep inelastic scattering \cite{ESW96}.  It was optimistically imagined that $Q_0$ might be as small as 1\,GeV and so CEBAF was designed to reach 4\,GeV.

In the ten years following achievement of design capacity, numerous fascinating experimental results were obtained at JLab, including an empirical demonstration that the distribution of charge and magnetization within the proton are completely different \cite{Jones:1999rz,Gayou:2001qd,Gayou:2001qt,Punjabi:2005wq,Puckett:2010ac,Puckett:2011xg}, \emph{viz}.\
\begin{equation}
\mu_p G_E^p(Q^2)/G_M^p(Q^2)\neq 1\,.
\end{equation}
This fact, which overturned a longstanding particle physics paradigm, along with a range of related observations, can be explained by the presence of strong, nonpointlike, electromagnetically-active scalar and axial-vector diquark correlations within the nucleon \cite{Eichmann:2008ef,Cloet:2008re,Chang:2011tx,Cloet:2011qu,Wilson:2011aa,Cloet:2013gva,%
Cloet:2014rja,Segovia:2014aza}.  However, notwithstanding the breadth of JLab's programme and the excitement it has generated, no experiment has yet produced an unambiguous signal for the realisation of Eq.\,\eqref{QCDtest}, \emph{i.e}.\ as yet, we have no clear sign of parton model scaling and certainly no sign of scaling violations.  Partly owing to this but also with a vast array of new experimental tests of the Standard Model's strong interaction sector in mind \cite{Dudek:2012vr}, it was decided in 2004 that CEBAF should be upgraded so that it could deliver 12\,GeV electrons on targets.  That upgrade is now complete and commissioning beams are being sent to the experimental halls.

The JLab\,12 facility and, indeed, an array of modern accelerators worldwide, will confront a diverse range of scientific challenges.
In the foreseeable future, we will know the results of a search for hybrid and exotic hadrons, the discovery of which would force a dramatic reassessment of the distinction between matter and force fields in Nature.
Opportunities provided by new data on hadron elastic and transition form factors will be exploited, yielding insights into the infrared running of QCD's coupling and dressed-masses, revealing correlations that are key to hadron structure, exposing the facts or fallacies in contemporary descriptions of hadron structure, and seeking for verification of Eq.\,\eqref{QCDtest} in numerous processes.
Precision experimental studies of the valence-quark region will proceed, with the results being used to confront predictions from theoretical computations of distribution functions and distribution amplitudes -- computation is critical here because without it no amount of data will reveal anything about the theory underlying the phenomena of strong interaction physics.
In addition, the international community will seek and exploit opportunities to use precision-QCD as a probe for physics beyond the Standard Model.
Each of the pieces in this body of exploration, however, can be viewed as steps aimed toward understanding a single overarching puzzle within the Standard Model, \emph{viz}.\ what is confinement and how is it related to DCSB -- the origin of the vast bulk of visible mass.

In order to match the rate at which progress is being made experimentally and, better, to guide and enhance that programme, flexible, responsive theoretical approaches are needed: methods that are capable both of rapidly providing an intuitive understanding of complex problems and illuminating a path toward answers and new discoveries.  In this milieu, notwithstanding its steady progress toward results with input parameters that approximate the real world, the numerical simulation of lattice-regularised QCD will not suffice. Approaches formulated in the continuum and inspired by, based upon, or connected directly with QCD are necessary.  Prominent amongst such tools are QCD Sum Rules \cite{Nielsen:2009uh} and Dyson-Schwinger equations (DSEs) \cite{Cloet:2013jya}, both of which were canvassed at this Workshop.

\section{Quantum Chromodynamics}
QCD is a quantum gauge field theory based on the group $SU(3)$ \cite{Marciano:1979wa}; and following around forty years of study, there is no confirmed breakdown over an enormous energy domain: $0 < E < 8\,$TeV.  Consequently, QCD is plausibly the only known instance of a quantum field theory that can rigorously be defined nonperturbatively \cite{millennium:2006}, in which case it is truly a field theory \emph{not} \emph{merely} an effective field theory.  The possibility that QCD might be rigorously well defined is one of its deepest fascinations.  In that case, QCD could stand alone as an archetype -- the only internally consistent quantum field theory which is defined at all energy scales.  This is a remarkable possibility with wide-ranging consequences and opportunities, \emph{e}.\emph{g}. it means that QCD-like theories provide a viable paradigm for extending the Standard Model to greater scales than those already probed.  Contemporary research in this direction is typified by the notion of extended technicolour \cite{Andersen:2011yj,Sannino:2013wla}, in which electroweak symmetry breaks via a fermion bilinear operator in a strongly-interacting non-Abelian theory; and the Higgs sector of the Standard Model becomes an effective description of a more fundamental fermionic theory, similar to the Ginzburg-Landau theory of superconductivity.

\subsection{Confinement}
\label{secConfinement}
This notion has already been mentioned.  However, in order to consider the concept further it is actually crucial to \emph{define} the subject.  That problem is canvassed in Sec.\,2.2 of Ref.\,\cite{Cloet:2013jya}, which explains that the potential between infinitely-heavy quarks measured in numerical simulations of quenched lattice-regularised QCD -- the so-called static potential \cite{Wilson:1974sk} -- is \emph{irrelevant} to the question of confinement in our Universe, in which light quarks are ubiquitous and the pion is unnaturally light.  This is because light-particle creation and annihilation effects are essentially nonperturbative in QCD and so it is impossible in principle to compute a quantum mechanical potential between two light quarks \cite{Bali:2005fu,Chang:2009ae}.  This means there is no flux tube in a Universe with light quarks and consequently that the flux tube is not the correct paradigm for confinement.

An alternative perspective associates confinement with dramatic, dynamically-driven changes in the analytic structure of QCD's propagators and vertices.  In this realisation, confinement is a dynamical process.  In fact, as will subsequently be explained, contemporary theory predicts that both quarks and gluons acquire running mass distributions in QCD, which are large at infrared momenta (see, \emph{e}.\emph{g}.\ Refs.\,\cite{Bhagwat:2003vw,Bowman:2005vx,Bhagwat:2006tu,Boucaud:2011ug,Ayala:2012pb,Binosi:2014aea}).  The generation of these masses leads to the emergence of a length-scale $\varsigma \approx 0.5\,$fm, whose existence and magnitude is evident in all existing studies of dressed-gluon and -quark propagators and which characterizes a dramatic change in their analytic structure.  In models based on such features \cite{Stingl:1994nk}, once a gluon or quark is produced, it begins to propagate in spacetime; but after each ``step'' of length $\varsigma$, on average, an interaction occurs so that the parton loses its identity, sharing it with others.  Finally a cloud of partons is produced, which coalesces into colour-singlet final states.  Such pictures of parton propagation, hadronisation and confinement can be tested in experiments at modern and planned facilities.

\begin{figure}[tbp]
\begin{minipage}[t]{\textwidth}
\begin{minipage}{0.5\textwidth}
\centerline{\includegraphics[clip,width=0.45\textwidth,angle=-90]{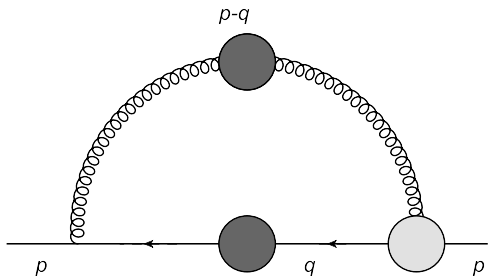}}
\end{minipage}
\begin{minipage}{0.5\textwidth}{\small
\caption{\small
\label{FigSigma} Dressed-quark self energy, which is the dynamical content of QCD's most basic fermion gap equation.  The kernel is composed from the dressed-gluon propagator (spring with dark circle) and the dressed-quark-gluon vertex (light-circle).  The equation is nonlinear owing to the appearance of the dressed-quark propagator (line with dark circle).  This image encodes every valid Feynman diagram relevant to the quark dressing process.  (Momentum flows from right-to-left.)}}
\end{minipage}
\end{minipage}
\end{figure}

\subsection{Dynamical chiral symmetry breaking}
Whilst the nature and realisation of confinement in empirical QCD is still being explored, DCSB; namely, the generation of \emph{mass} \emph{from nothing}, is a theoretically-established nonperturbative feature of QCD.  It is important to insist on the term ``dynamical,'' as distinct from spontaneous, because nothing is added to QCD in order to effect this remarkable outcome and there is no simple change of variables in the QCD action that will make it apparent.  Instead, through the act of quantising the classical chromodynamics of massless gluons and quarks, a large mass-scale is generated.  DCSB is the most important mass generating mechanism for visible matter in the Universe, being responsible for approximately $98$\% of the proton's mass.

\begin{figure}[t]
\begin{minipage}[t]{\textwidth}
\begin{minipage}{0.5\textwidth}
\centerline{\includegraphics[clip,width=0.9\textwidth]{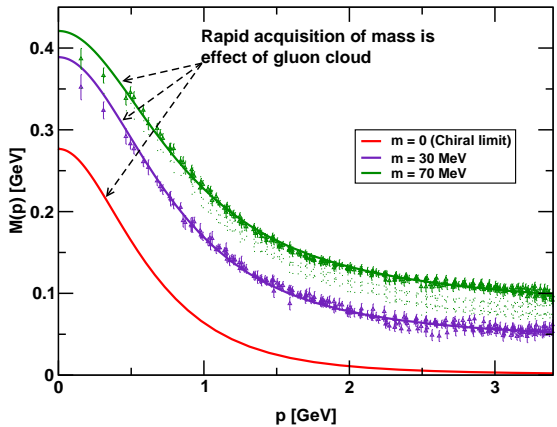}}
\end{minipage}
\begin{minipage}{0.5\textwidth}{\small
\caption{\label{gluoncloud} \small
Dressed-quark mass function, $M(p)$ in Eq.\,(\protect\ref{SgeneralN}): \emph{solid curves} -- DSE results, explained in Refs.\,\protect\cite{Bhagwat:2003vw,Bhagwat:2006tu}, ``data'' -- numerical simulations of lattice-regularised QCD \protect\cite{Bowman:2005vx}.  (NB.\ $m=70\,$MeV is the uppermost curve and current-quark mass decreases from top to bottom.)  The current-quark of perturbative QCD evolves into a constituent-quark as its momentum becomes smaller.  The constituent-quark mass arises from a cloud of low-momentum gluons attaching themselves to the current-quark.  This is DCSB: an essentially nonperturbative effect that generates a quark \emph{mass} \emph{from nothing}; namely, it occurs even in the chiral limit.
}\vspace*{5ex}}
\end{minipage}
\end{minipage}
\end{figure}

A fundamental expression of DCSB is the behaviour of the quark mass-function, $M(p)$, which is a basic element in the dressed-quark propagator
\begin{equation}
\label{SgeneralN}
S(p) = 
1/[i\gamma\cdot p A(p^2) + B(p^2)] = Z(p^2)/[i\gamma\cdot p + M(p^2)]\,,
\end{equation}
which may be obtained as a solution to QCD's most basic fermion gap equation (see Fig.\,\ref{FigSigma}).  The highly nontrivial behaviour of the mass function, depicted in Fig.\,\ref{gluoncloud}, arises primarily because a dense cloud of gluons comes to clothe a low-momentum quark; and explains how an almost-massless parton-like quark at high energies transforms, at low energies, into a constituent-like quark, which possesses an effective ``spectrum mass'' $M_D \sim 350\,$MeV.  Consequently, the proton's mass is two orders-of-magnitude larger than the sum of the current-masses of its three valence-quarks.

\subsection{Whence the mass?}
One might ask just how the self-energy depicted in Fig.\,\ref{FigSigma} is capable of generating \emph{mass from nothing}, \emph{viz}.\ the $m=0$ curve in Fig.\,\ref{gluoncloud}, which cannot arise in the classical theory.  The answer lies in the fact that Fig.\,\ref{FigSigma} is a deceptively simply picture.  It actually corresponds to a countable infinity of diagrams, all of which can potentially contribute.  To provide a context, quantum electrodynamics, an Abelian gauge theory, has 12\,672 diagrams at order $\alpha^5$ in the computation of the electron's anomalous magnetic moment \cite{Aoyama:2012wj}.  Owing to its foundation in the non-Abelian group $SU(3)$, the analogous perturbative computation of a quark's anomalous chromomagnetic moment has many more diagrams at this order in the strong coupling.  The number of diagrams represented by the self energy in Fig.\,\ref{FigSigma} grows equally rapidly, \emph{i}.\emph{e}.\ combinatorially with the number of propagators and vertices used at a given order.  Indeed, proceeding systematically, a computer will very quickly generate the first diagram in which the number of loops is so great that it is simply impossible to calculate in perturbation theory: impossible in the sense that we don't yet have the mathematical capacity to solve the problem.

Each of the diagrams which contributes to $M(p^2)$ in a weak-coupling expansion of Fig.\,\ref{FigSigma} is multiplied by the current-quark mass, $\hat m$.  Plainly, any finite sum of diagrams must therefore vanish as $\hat m\to 0$.  However, with \emph{infinitely many} diagrams the situation might be very different: one has ``$ 0 \times \infty$,'' a product whose limiting value is contingent upon the cumulative magnitude of each term in the sum.  Consider therefore the behaviour of $M(p^2)$ at large $p^2$.  QCD is asymptotically free \cite{Politzer:2005kc,Gross:2005kv,Wilczek:2005az}.  Hence, on this domain, each of the regularised loop diagrams must individually evaluate to a small number whose value depends on just how large is the coupling.  It will not be surprising, therefore, to learn that for a monotonically-decreasing running-coupling, $\alpha_S(k^2)$, there is a critical value of $\alpha_S(0)$ above which the magnitude of the sum of infinitely many diagrams is sufficient to balance the linear decrease of $\hat m\to 0$, so that the answer is nonzero and finite in this limit, \emph{viz}.,
\begin{equation}
\exists\, \alpha_S^c(0) \; |\; \forall \alpha_S(0) >\alpha_S^c(0),
M_0(p^2):= \lim_{\hat m\to 0} M(p^2;\hat m) \neq 0\,.
\end{equation}
The internal consistency of QCD appears to guarantee that the limit is always finite.  (The case of Abelian theories is more complicated \cite{Roberts:1994dr} because they are not asymptotically free.)

\subsection{Enigma of mass}
The pion is Nature's lightest hadron.  In fact, it is peculiarly light, with a mass just one-fifth of that which quantum mechanics would lead one to expect.  This remarkable feature has its origin in DCSB.  In quantum field theory the pion's structure is described by a Bethe-Salpeter amplitude (here $k$ is the relative momentum between the  valence-quark and -antiquark constituents, and $P$ is their total momentum):
\begin{equation}
\Gamma_{\pi}(k;P) = \gamma_5 \left[
i E_{\pi}(k;P) + \gamma\cdot P F_{\pi}(k;P)  + \gamma\cdot k \, G_{\pi}(k;P) - \sigma_{\mu\nu} k_\mu P_\nu H_{\pi}(k;P)
\right],
\label{genGpi}
\end{equation}
which is simply related to an object that would be its Schr\"odinger wave function if a nonrelativistic limit were appropriate.  In QCD if, and only if, chiral symmetry is dynamically broken, then for $\hat m=0$ \cite{Maris:1997hd,Qin:2014vya}:
\begin{equation}
\label{gtrE}
f_\pi E_\pi(k;0) = B(k^2)\,,
\end{equation}
where the right-hand-side is a scalar function in the dressed-quark propagator, Eq.\,\eqref{SgeneralN}.  This identity is miraculous.  It means that the two-body problem is solved, almost completely, without lifting a finger, once the solution to the one body problem is known.  Eq.\,\eqref{gtrE} is a quark-level Goldberger-Treiman relation.  It is also the most basic expression of Goldstone's theorem in QCD, \emph{viz}.\\[-3ex]

\centerline{\parbox{0.95\textwidth}{\flushleft \emph{Goldstone's theorem is fundamentally an expression of equivalence between the one-body problem and the two-body problem in QCD's colour-singlet pseudoscalar channel.}}}

\medskip

\hspace*{-\parindent}Eq.\,\eqref{gtrE} emphasises that Goldstone's theorem has a pointwise expression in QCD; and, furthermore, that pion properties are an almost direct measure of the mass function depicted in Fig.\,\ref{gluoncloud}.  Thus, enigmatically, properties of the (nearly-)massless pion are the cleanest expression of the mechanism that is responsible for almost all the visible mass in the Universe.  Plainly, DCSB has a very deep and far-reaching impact on physics within the Standard Model.

\subsection{Gluon cannibalism}
It is not just the propagation of quarks that is affected by strong interactions in QCD.  The propagation of gluons, too, is described by a gap equation; and its solution shows that gluons are cannibals: they are a particle species whose members become massive by eating each other!  The associated gluon mass function, $m_g(k^2)$, is monotonically decreasing with increasing $k^2$ and recent work \cite{Binosi:2014aea} has established that
\begin{equation}
\label{gluonmassEq}
m_g(k^2=0) \approx 0.5\,{\rm GeV}.
\end{equation}
The value of the mass-scale in Eq.\,\eqref{gluonmassEq} is \emph{natural} in the sense that it is commensurate with but larger than the value of the dressed light-quark mass function at far infrared momenta: $M(0)\approx 0.3\,$GeV (see Fig.\,\ref{gluoncloud}). Moreover, the mass term appears in the transverse part of the gluon propagator, hence gauge-invariance is not tampered with; and the mass function falls as $1/k^2$ for $k^2\gg m_g(0)$ (up to logarithmic corrections), so the gluon mass is invisible in perturbative applications of QCD.

Gauge boson cannibalism presents a new physics frontier within the Standard Model. Asymptotic freedom means that the ultraviolet behaviour of QCD is controllable.  At the other extreme, dynamically generated masses for gluons and quarks entail that QCD creates its own infrared cutoffs.  Together, these effects eliminate both the infrared and ultraviolet problems that typically plague quantum field theories and thereby make reasonable the hope that QCD is nonperturbatively well defined.

The dynamical generation of gluon and quark masses provides a basis for understanding the notion of a maximum wavelength for gluons and quarks in QCD \cite{Brodsky:2008be}.  Indeed, given the magnitudes of the gluon and quark mass-scales, it is apparent that field modes with wavelengths $\lambda > \varsigma \approx 2/m_g(0) \approx 0.5\,$fm decouple from the dynamics.  They are screened in the sense described in Sec.\,\ref{secConfinement}.  This is just one consequence of the appearance of a dynamically generated gluon mass-scale.

There are many more.  For example, the exceptionally light pion degree-of-freedom becomes dominant in QCD at those length-scales above which dressed-gluons and -quarks decouple from the theory owing to the large magnitudes of their dynamically generated masses.  It is therefore conceivable that Gribov copies have no measurable impact on observables within the Standard Model because they affect only those gluonic modes whose wavelengths lie in the far infrared; and such modes are dynamically screened, by an exponential damping factor $\sim \exp(-\lambda/\varsigma)$, so that their role in hadron physics is superseded by the dynamics of light-hadrons.  This conjecture is consistent with the insensitivity to Gribov copies of the dressed-gluon and -quark two-point Schwinger functions observed in numerical simulations of QCD on fine lattices \cite{Bowman:2002fe,Zhang:2004gv}.  Another plausible conjecture is that dynamical generation of an infrared gluon mass-scale leads to saturation of the gluon parton distribution function at small Bjorken-$x$ within hadrons. The possible emergence of this phenomenon stirs great scientific interest and curiosity.  It is a key motivation in plans to construct an electron ion collider (EIC) that would be capable of producing a precise understanding of collective behaviour amongst gluons \cite{Accardi:2012qut}.

\section{Continuum-QCD and ab initio predictions of hadron observables}
\label{secAbInitio}
Within contemporary hadron physics there are two common methods for determining the momentum-dependence of the interaction between quarks: the top-down approach, which works toward an \textit{ab initio} computation of the interaction via direct analysis of the gauge-sector gap equations; and the bottom-up scheme, which aims to infer the interaction by fitting data within a well-defined truncation of those equations in the matter sector that are relevant to bound-state properties.  These two approaches have recently been united \cite{Binosi:2014aea} by a demonstration that the renormalisation-group-invariant (RGI) running-interaction predicted by contemporary analyses of QCD's gauge sector coincides with that required in order to describe ground-state hadron observables using a nonperturbative truncation of QCD's Dyson-Schwinger equations in the matter sector, \emph{i.e}.\ the DCSB-improved (DSE-DB) kernel elucidated in Refs.\,\cite{Chang:2009zb,Chang:2010hb,Chang:2011ei}.

\begin{figure}[t]
\begin{minipage}[t]{\textwidth}
\begin{minipage}{0.48\textwidth}
\centerline{\includegraphics[clip,width=0.9\textwidth]{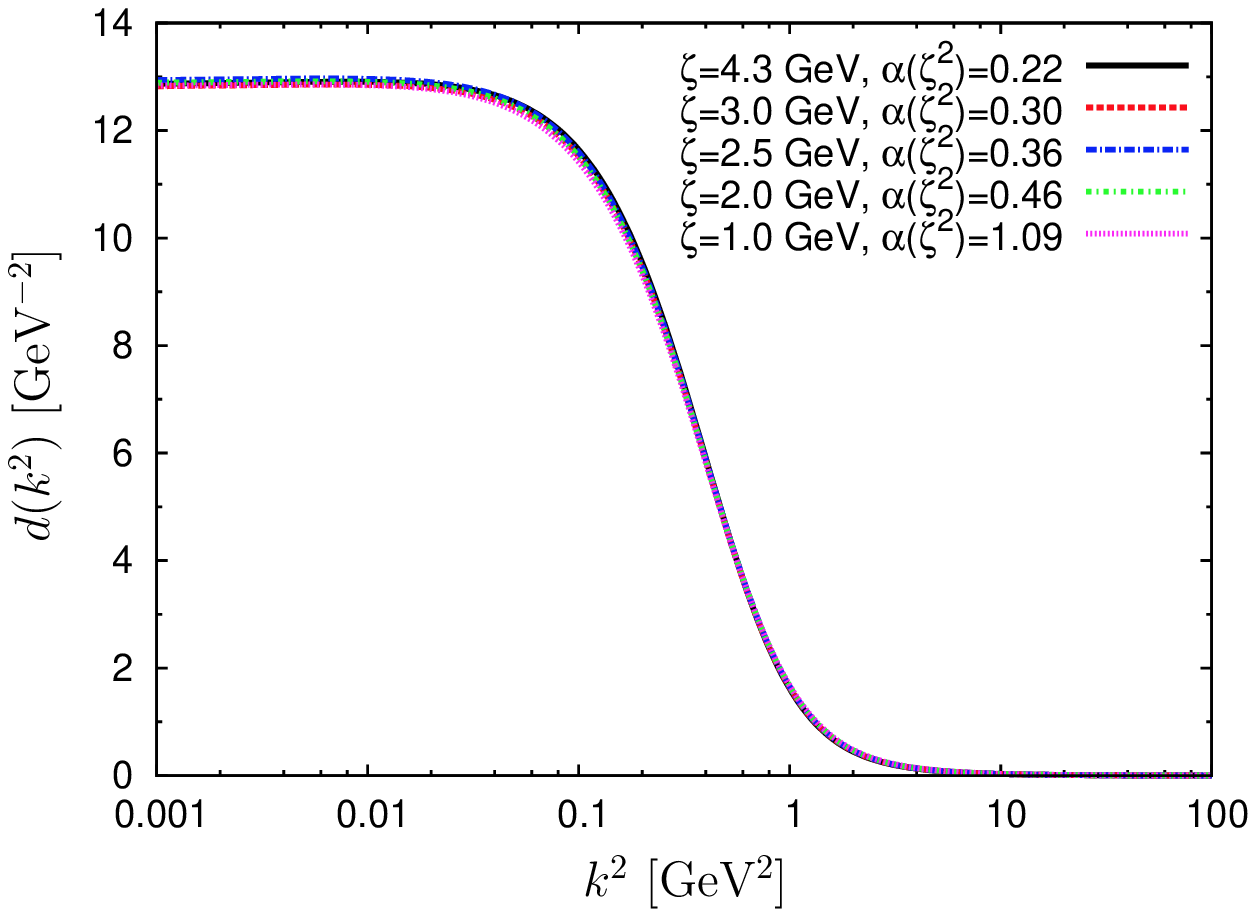}}
\end{minipage}
\begin{minipage}{0.02\textwidth}
\hspace*{-0.2em}\mbox{\LARGE \textbf{$\Rightarrow$}}
\end{minipage}
\begin{minipage}{0.48\textwidth}
\centerline{\includegraphics[clip,width=0.9\textwidth]{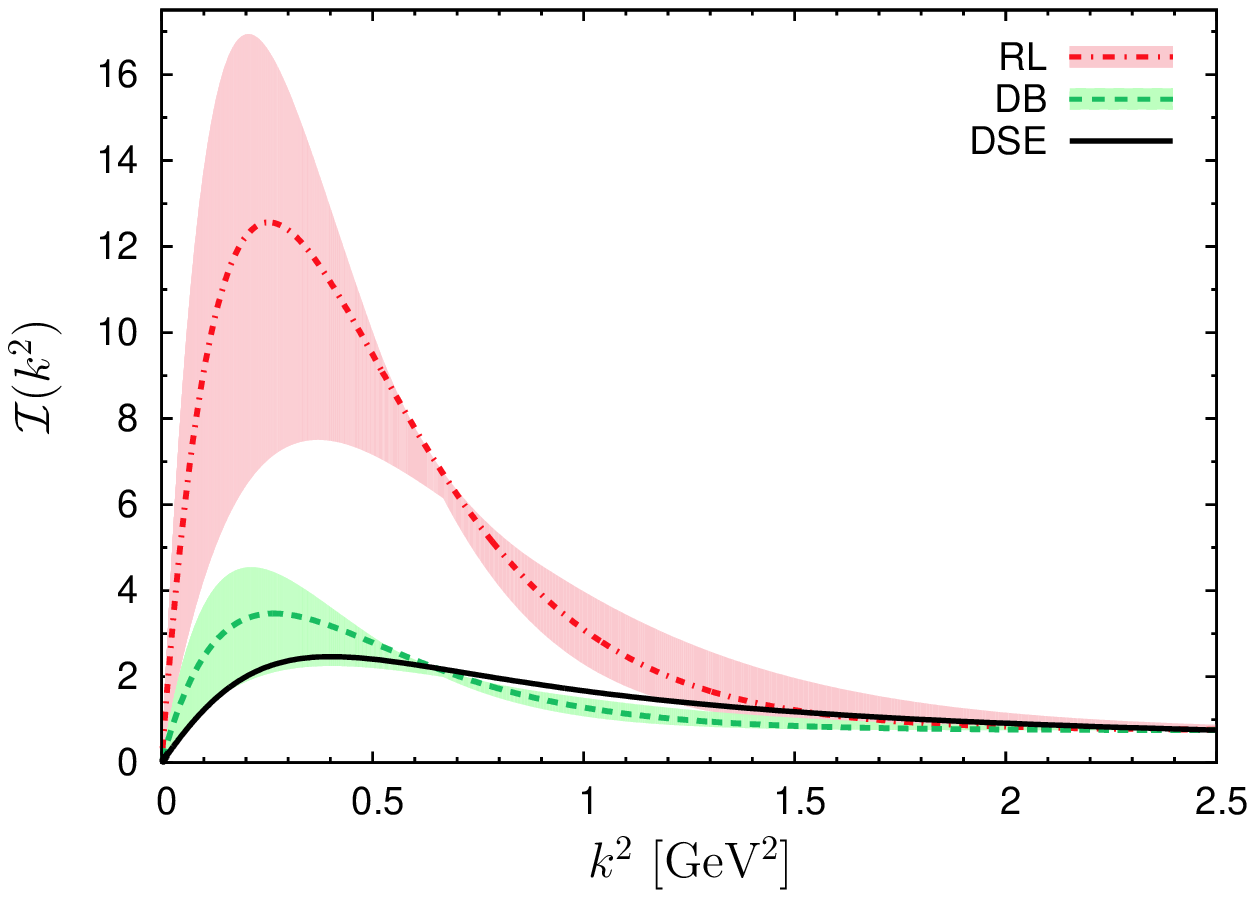}}
\end{minipage}
\end{minipage}
\caption{\label{figInteraction} \small
\emph{Left panel} -- RGI running interaction strength computed via a combination of DSE- and lattice-QCD analyses, as explained in  Ref.\,\cite{Aguilar:2009nf}.  The function obtained with five different values of the renormalisation point is depicted in order to highlight that the result is RGI.  The interaction is characterized by a value $\alpha_s(0) \approx 0.9\, \pi$ and the gluon mass-scale in Eq.\,\eqref{gluonmassEq}.
\emph{Right panel} -- Comparison between top-down results for the gauge-sector interaction (derived from the left-panel) with those obtained using the bottom-up approach based on hadron physics observables.  \emph{Solid curve} -- top-down result for the RGI running interaction; \emph{dot-dashed curve} within \emph{pale-red band} -- bottom-up result obtained in the RL truncation; and \emph{dashed curve} within \emph{pale-green band} -- advanced bottom-up result obtained in the DB truncation.  The bands denote the domain of uncertainty in the bottom-up determinations of the interaction.  All curves are identical on the perturbative domain: $k^2>2.5\,$GeV$^2$.
}
\end{figure}

The unification is illustrated in Fig.\,\ref{figInteraction}: the right panel presents a comparison between the top-down RGI interaction (solid-black curve, derived from the left panel) and the DB-truncation bottom-up interaction (green band containing dashed curve).  Plainly, the interaction predicted by modern analyses of QCD's gauge sector is in near precise agreement with that required for a veracious description of measurable hadron properties using the most sophisticated matter-sector gap and Bethe-Salpeter kernels available today.  This is a remarkable result, given that there had previously been no serious attempt at communication between practitioners from the top-down and bottom-up hemispheres of continuum-QCD.  It bridges a gap that had lain between nonperturbative continuum-QCD and the \emph{ab initio} prediction of bound-state properties.

A comparison between the top-down prediction and that inferred using the simple DSE-RL kernel (red band containing dot-dashed curve) is also important.  One observes that the DSE-RL result has the correct shape but is too large in the infrared.  This is readily explained \cite{Binosi:2014aea}; and it follows that whilst the RL truncation supplies a useful computational link between QCD's gauge sector and measurable hadron properties, the model interaction it delivers is \emph{not a pointwise-accurate} representation of ghost-gluon dynamics.  Notwithstanding this, it remains true that the judicious use of RL truncation can yield reliable predictions for a known range of hadron observables, with an error that may be estimated and whose origin is understood.

\section{A Physical Vacuum}
\label{secVacuum}
\begin{figure}[t]
\begin{minipage}[t]{\textwidth}
\begin{minipage}{0.45\textwidth}
\centerline{\includegraphics[clip,width=0.9\textwidth]{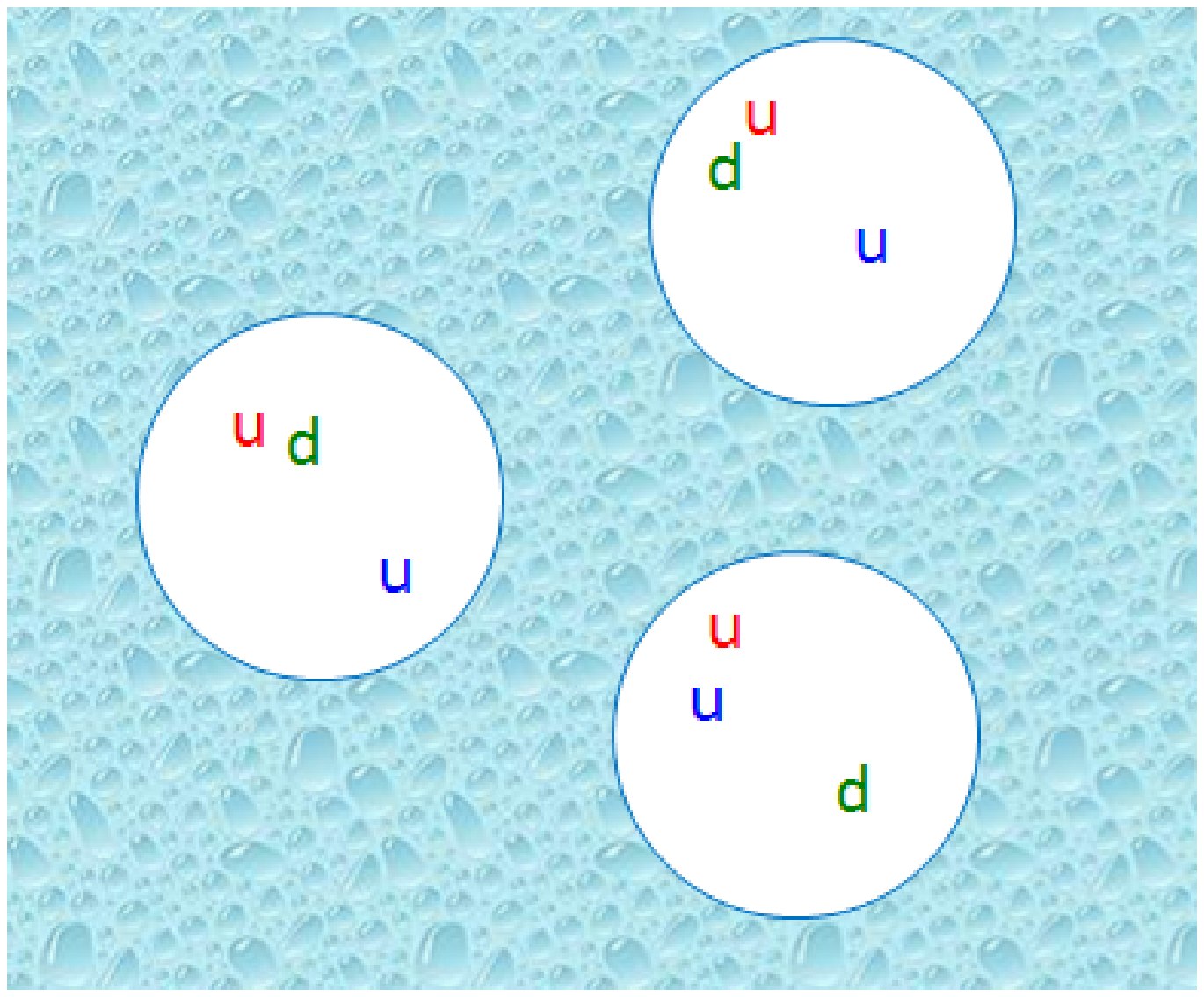}}
\end{minipage}
\begin{minipage}{0.10\textwidth}
\hspace*{2em}\mbox{\LARGE \textbf{$\Rightarrow$}}
\end{minipage}
\begin{minipage}{0.45\textwidth}
\centerline{\includegraphics[clip,width=0.9\textwidth]{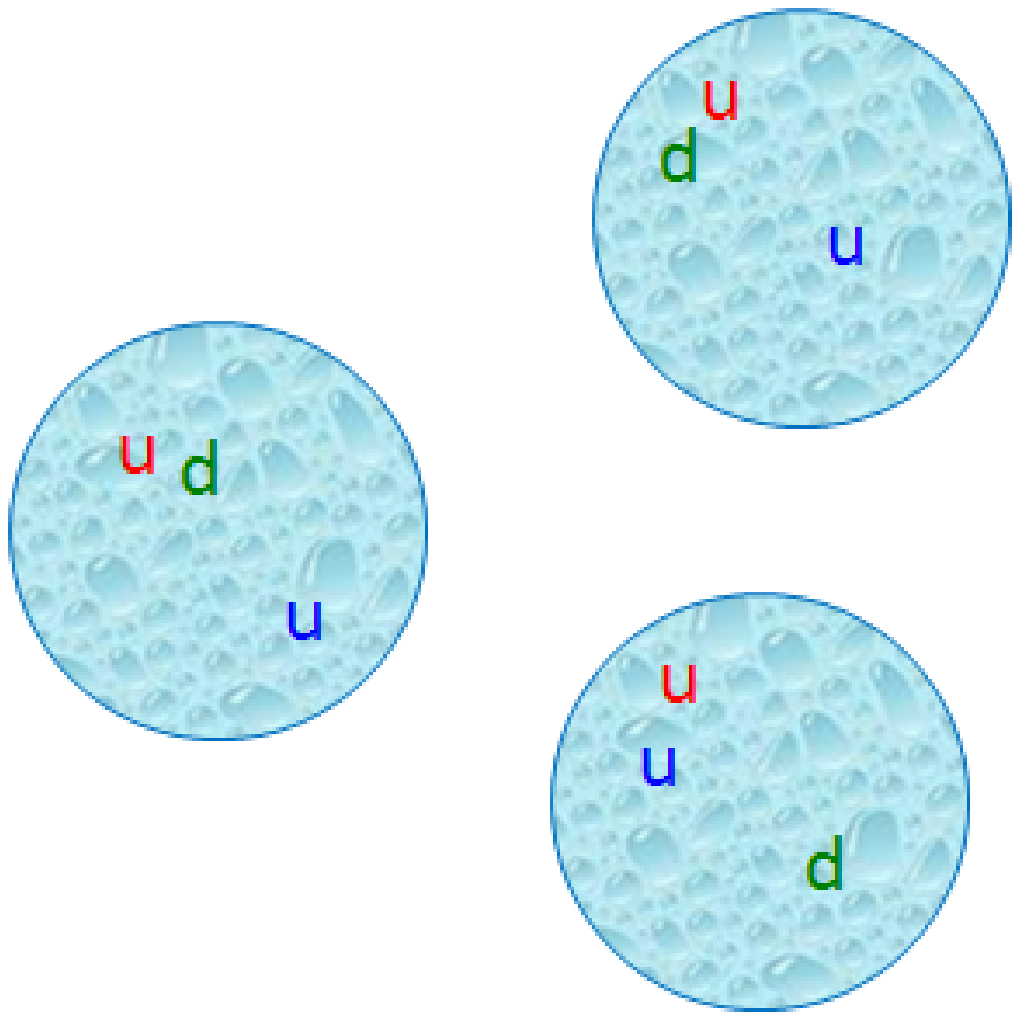}}
\end{minipage}
\end{minipage}
\caption{\label{vacuum} \small
\emph{Left panel} -- The QCD vacuum was historically imagined to be a ``frothing sea,'' with hadrons being merely bubbles of foam containing nothing but quarks and gluons interacting perturbatively throughout.  Only near the bubble's boundary did some sort of strong interaction occur, enforcing quark and gluon confinement.
\emph{Right panel} -- A modern view \cite{Brodsky:2009zd,Brodsky:2010xf,Chang:2011mu,Brodsky:2012ku} flips the picture completely.  The space between hadrons is ``empty,'' except for the perturbative quantum fluctuations that characterise all quantum field theories.  The interior of hadrons, however, is extremely complex, with nonperturbative dynamics dominating in $\sim 98$\% of the interior: the so-called condensates are spacetime-dependent and confined within the hadrons.
}
\end{figure}

As the preceding discussion indicates, DCSB is a crucial emergent feature of the Standard Model.  It is very clearly expressed in the dressed-quark mass function of Fig.\,\ref{gluoncloud}.  However, this understanding is relatively recent.  DCSB was historically conflated with the existence of a spacetime-independent quark-antiquark condensate, $\langle \bar q q\rangle$, that permeates the Universe.  This notion was born with the introduction of QCD sum rules as a theoretical artifice to estimate nonperturbative strong-interaction matrix elements \cite{Shifman:1978bx,Leinweber:1995fn} and is typically tied to a belief that the QCD vacuum is characterized by numerous distinct, spacetime-independent condensates, as illustrated in the left panel of Fig.\,\ref{vacuum}.

This belief is harmless unless one imagines that the theory of gravity is understood well enough so that it may reliably be coupled to quantum field theory.  Subscribers to this view argue \cite{Turner:2001yu,Bass:2011zz} that the energy-density of the Universe must receive a contribution from such vacuum condensates and that the only possible covariant form for the energy of the (quantum) vacuum, \emph{viz}.
\begin{equation}
T_{\mu\nu}^{\rm VAC} = \rho_{\rm VAC}\, \delta_{\mu\nu}\,,
\end{equation}
is mathematically equivalent to the cosmological constant.  From this perspective, the quantum vacuum is \cite{Turner:2001yu} ``\ldots a perfect fluid and precisely spatially uniform \ldots'' so that ``Vacuum energy is almost the perfect candidate for dark energy.''  Now, if the ground state of QCD is really expressed in a nonzero spacetime-independent expectation value $\langle\bar q q\rangle$, then the energy difference between the symmetric and broken phases is roughly $M_{\rm QCD} \sim 0.3\,$GeV, as indicated by Fig.\,\ref{gluoncloud}.  One obtains therefrom:
\begin{equation}
\rho_\Lambda^{\rm QCD} = 10^{46} \rho_\Lambda^{\rm obs},
\end{equation}
\emph{i.e}.\ the contribution from the QCD vacuum to the energy density associated with the cosmological constant exceeds the observed value by forty-six orders-of-magnitude.  In fact, the discrepancy is far greater if the Higgs vacuum expectation value is treated in a similar manner.

This mismatch has been called \cite{Zee:2008zza} ``\ldots one of the gravest puzzles of theoretical physics.''  However, it vanishes if one discards the notion that condensates have a physical existence, which is independent of the hadrons that express QCD's asymptotically realisable degrees of freedom \cite{Brodsky:2009zd}; namely, if one accepts that such condensates are merely mass-dimensioned parameters in one or another theoretical computation and truncation scheme.  This appears mandatory in a confining theory \cite{Chang:2011mu,Brodsky:2010xf,Brodsky:2012ku}, a perspective one may embed in a broader context by considering just what is observable in quantum field theory \cite{Weinberg:1978kz}: ``\ldots although individual quantum field theories have of course a good deal of content, quantum field theory itself has no content beyond analyticity, unitarity, cluster decomposition and symmetry.''  If QCD is a confining theory, then the principle of cluster decomposition is only realised for colour singlet states \cite{Krein:1990sf} and all observable consequences of the theory, including its ground state, can be expressed via a hadronic basis.  This is quark-hadron duality.

The new hypothesis \cite{Brodsky:2009zd,Brodsky:2010xf,Chang:2011mu,Brodsky:2012ku} can therefore be stated succinctly as follows: ``If quark-hadron duality is a reality in QCD, then condensates, those quantities that have commonly been viewed as constant empirical mass-scales that fill all spacetime, are instead wholly contained within hadrons, \emph{i.e}.\ they are a property of hadrons themselves and expressed, for example, in their Bethe-Salpeter or light-front wave functions.''  This view is depicted in the right panel of Fig.\,\ref{vacuum} and canvassed fully in Sec.\,4 of Ref.\,\cite{Cloet:2013jya}.  It presents the reasonable perspective that the understanding of hadrons requires that one explain what lies within those hadrons in contrast to the historical alternative, which suggested that hadrons could only be understood by explaining the properties of the vast spacetime domains that contain no hadrons at all.

The shift in perspective highlighted by the right panel of Fig.\,\ref{vacuum} does not undermine the utility of the QCD sum rules approach to the estimation of hadron observables.  Instead, it tames the condensates so that they return to being merely mass-dimensioned parameters in a valuable computation scheme.  Its implications are, however, significant and wide-ranging.  For example, in connection with the cosmological constant, putting QCD condensates back into hadrons reduces the mismatch between experiment and theory by a factor of $10^{46}$.  Furthermore, if technicolour-like theories \cite{Andersen:2011yj,Sannino:2013wla} are the correct scheme for explaining electroweak symmetry breaking, then the impact of the notion of in-hadron condensates is far greater still.

\section{Hadron interiors}
\label{sec-1}
Since the advent of the parton model and the first deep inelastic scattering experiments there has been a determined effort to deduce the parton distribution functions (PDFs) of the most stable hadrons \cite{Holt:2010vj}.  The behaviour of such distributions on the valence domain (Bjorken-$x> 0.5$) is of particular interest because this domain is definitive of hadrons, \emph{e.g}.\ quark content on the valence domain is how one distinguishes between a neutron and a proton: a neutron possesses one valence $u$-quark plus two valence $d$-quarks whereas the proton possesses two valence $u$-quarks plus one valence $d$-quark.  Indeed, all Poincar\'e-invariant properties of a hadron: baryon number, charge, total spin, \emph{etc}., are determined by the PDFs which dominate on the valence domain.  Moreover, via QCD evolution \cite{Dokshitzer:1977,Gribov:1972,Lipatov:1974qm,Altarelli:1977}, PDFs on the valence-quark domain determine backgrounds at the large hadron collider.  There are also other questions, \emph{e.g}.\ regarding flavour content of a hadron's sea and whether that sea possesses an intrinsic component \cite{Brodsky:1980pb,Signal:1987gz}.  The answers to all these questions are essentially nonperturbative properties of QCD.

Recognising the significance of the valence domain, a new generation of experiments, focused on $x\gtrsim 0.5$, is planned at JLab, and under examination in connection with Drell-Yan studies at Fermilab and a possible EIC.  Consideration is also being given to experiments aimed at measuring parton distribution functions in mesons at J-PARC.  Furthermore, at FAIR it would be possible to directly measure the Drell-Yan process from high-$x$ antiquarks in the antiproton annihilating with quarks in the proton.  

A concentration on such measurements requires theory to move beyond merely parametrising distribution functions and amplitudes.  Computation within QCD-connected frameworks becomes critical because without it, no amount of data will reveal anything about the theory underlying strong interaction phenomena.  This is made clear by the example of the pion's valence-quark PDF, $u_v^\pi(x)$, in connection with which a failure of QCD was suggested following a leading-order analysis of $\pi N$ Drell-Yan measurements \cite{Conway:1989fs}.  As explained in Ref.\,\cite{Holt:2010vj}, this confusion was fostered by the application of a diverse range of models.  On the other hand, a series of QCD-connected calculations  \cite{Hecht:2000xa,Aicher:2010cb,Nguyen:2011jy,Chang:2014lva} subsequently established that the leading-order analysis was misleading, so that $u_v^\pi(x)$ may now be seen as a success for the unification of nonperturbative and perturbative studies in QCD.

A framework that provides access to the pion's valence-quark PDF can also be employed to compute its PDAs.  For example, the pion's leading-twist two-particle PDA is given by the following projection of the pion's Bethe-Salpeter wave function onto the light-front \cite{Chang:2013pq}
\begin{equation}
f_\pi\, \varphi_\pi(x;\zeta) = {\rm tr}_{\rm CD}
Z_2 \! \int_{dq}^\Lambda \!\!
\delta(n\cdot q_\eta - x \,n\cdot P) \,\gamma_5\gamma\cdot n\,
S(q_\eta)\Gamma(q;P)S(q_{\bar\eta})\,,
\label{pionPDA}
\end{equation}
where: $f_\pi$ is the pion's leptonic decay constant; $\int_{dq}^\Lambda$ is a Poincar\'e-invariant regularization of the four-dimensional integral, with $\Lambda$ the ultraviolet regularization mass-scale; $Z_{2}(\zeta,\Lambda)$ is the quark wave-function renormalisation constant, with $\zeta$ the renormalisation scale; $n$ is a light-like four-vector, $n^2=0$; and $P$ is the pion's four-momentum, $P^2=-m_\pi^2$ and $n\cdot P = -m_\pi$, with $m_\pi$ being the pion's mass.

\begin{figure}[t]
\begin{minipage}{\textwidth}
\begin{minipage}{0.5\textwidth}
\centerline{\includegraphics[width=0.95\textwidth,clip]{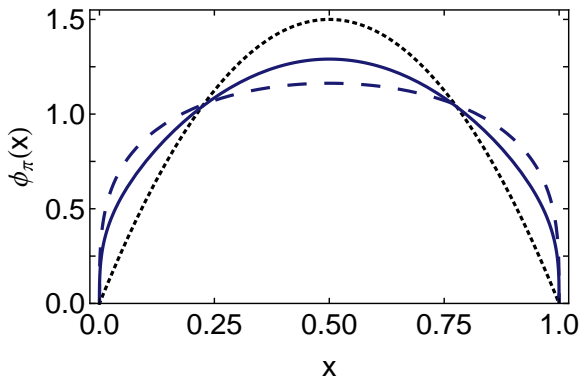}}
\end{minipage}
\begin{minipage}{0.5\textwidth}
\caption{\label{FigpionPDA} \small
Twist-two pion PDA computed using two very different DSE truncations at a scale $\zeta=2\,$GeV. \emph{Dashed curve} -- rainbow-ladder (RL), the leading-order in a systematic, sym\-metry-preserving scheme \protect\cite{Munczek:1994zz,Bender:1996bb}; and \emph{solid curve} -- the most sophisticated kernel that is currently available; namely, a DCSB-improved (DB) kernel that incorporates nonperturbative effects generated by DCSB that are omitted in RL truncation and any stepwise improvement thereof \cite{Chang:2009zb,Chang:2010hb,Chang:2011ei}.  These results are consistent with contemporary lattice-QCD \cite{Cloet:2013tta,Segovia:2013eca,Shi:2014uwa}.  The dotted curve is $\varphi^{\rm asy}_\pi(x)=6 x(1-x)$, the result obtained in conformal QCD \cite{Efremov:1979qk,Lepage:1979zb}.}
\end{minipage}
\end{minipage}
\end{figure}

The amplitude in Eq.\,\eqref{pionPDA} has been computed using two very different truncations of QCD's DSEs \cite{Chang:2013pq}, with the result depicted in Fig.\,\ref{FigpionPDA}.  Both kernels agree: compared with the asymptotic form, which is valid when $\zeta$ is extremely large, there is a marked broadening of $\varphi_\pi(x;\zeta)$, which owes exclusively to DCSB.  This causal connection may be claimed because the PDA is computed at a low renormalisation scale in the chiral limit, whereat the quark mass function owes entirely to DCSB via Eq.\,\eqref{gtrE}.  Moreover, the dilation measures the rate at which a dressed-quark approaches the asymptotic bare-parton limit.  It can be verified empirically at JLab12, \emph{e.g}.\ in measurements of the pion's electromagnetic form factor, the ratio of the proton's electric and magnetic form factors, and the behaviour of the form factors that characterise transitions between the nucleon and its excited states.

A question of more than thirty-years standing can be answered using Fig.\,\ref{FigpionPDA}; namely, when does $\varphi^{\rm asy}_\pi(x)$ provide a good approximation to the pion PDA?  Plainly, not at $\zeta=2\,$GeV.  The ERBL evolution equation \cite{Efremov:1979qk,Lepage:1979zb} describes the $\zeta$-evolution of $\varphi_\pi(x;\zeta)$; and applied to $\varphi_\pi(x;\zeta)$ in Fig.\,\ref{FigpionPDA}, one finds \cite{Cloet:2013tta,Segovia:2013eca,Shi:2014uwa} that $\varphi^{\rm asy}_\pi(x)$ is a poor approximation to the true result even at $\zeta=200\,$GeV.  Thus at empirically accessible energy scales, the PDAs of ground-state hadrons are ``squat and fat''.  Evidence supporting this picture had long been accumulating \cite{Mikhailov:1986be,Petrov:1998kg,Braun:2006dg,Brodsky:2006uqa}; and the dilation is verified by simulations of lattice-QCD \cite{Cloet:2013tta,Segovia:2013eca,Shi:2014uwa}.

\section{Electromagnetic properties of hadrons}
\label{secEM}
It is now possible to add flesh to the bones of Eq.\,\eqref{QCDtest}.  In the theory of strong interactions, the cross-sections for many hard exclusive hadronic reactions can be expressed in terms of the PDAs of the hadrons involved.  An example is the pion's elastic electromagnetic form factor, for which the prediction can be stated succinctly  \cite{Farrar:1979aw,Efremov:1979qk,Lepage:1979zb,Lepage:1980fj}:
\begin{equation}
\label{pionUV}
\exists Q_0>\Lambda_{\rm QCD} \; |\;  Q^2 F_\pi(Q^2) \stackrel{Q^2 > Q_0^2}{\approx} 16 \pi \alpha_s(Q^2)  f_\pi^2 \mathpzc{w}_\varphi^2,
\quad
\mathpzc{w}_\varphi = \frac{1}{3} \int_0^1 dx\, \frac{1}{x} \varphi_\pi(x)\,,
\end{equation}
where
$
\alpha_s(Q^2) = 4 \pi/[\beta_0\,\ln(Q^2/\Lambda^2_{\rm QCD})],
$
with $\beta_0 = 11 - (2/3) n_f$ ($n_f$ is the number of active quark flavours), is the leading-order expression for the strong running coupling.  As noted in the Introduction, the value of $Q_0$ is not predicted by perturbative QCD.  Here $\Lambda_{\rm QCD} \sim 0.2\,$GeV is the natural mass-scale of QCD, whose dynamical generation through quantisation spoils the conformal invariance of the classical massless theory \cite{Collins:1976yq,Nielsen:1977sy,tarrach}: $\Lambda_{\rm QCD}$, whose value must be determined empirically within the Standard Model, sets the scale for the dynamically generated masses described above.

\begin{figure}[t]
\begin{minipage}[t]{\textwidth}
\begin{minipage}{0.5\textwidth}
\centerline{\includegraphics[clip,width=0.95\textwidth]{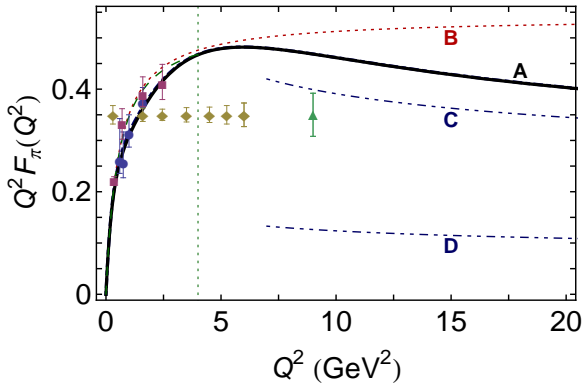}}
\end{minipage}
\begin{minipage}{0.5\textwidth}{\small
\caption{\label{figWPFpi} \small
$Q^2 F\pi(Q^2)$.  Solid curve (A) - Theoretical prediction \cite{Chang:2013nia}; dotted curve (B) -monopole form fitted to data;  dot-dot-dashed curve (C) - perturbative QCD (pQCD) prediction computed with the modern, dilated pion PDA described in Sec.\,\protect\ref{sec-1}; and dot-dot-dashed curve (D) - pQCD prediction computed with the asymptotic PDA, which had previously been used to guide expectations for the asymptotic behaviour of $Q^2 F\pi(Q^2)$.  The filled-circles and -squares represent existing JLab data \cite{Huber:2008id}; and the filled diamonds and triangle, whose normalisation is arbitrary, indicate the projected $Q^2$-reach and accuracy of forthcoming experiments \cite{E1206101,E1207105}.}}
\end{minipage}
\end{minipage}
\end{figure}

It was hoped that JLab would verify this fundamental Standard Model prediction, Eq.\,\eqref{pionUV}; and in 2001, seven years after commencing operations, the facility provided the first high-precision pion electroproduction data for $F_\pi(Q^2)$ between $Q^2$ values of 0.6 and 1.6\,GeV$^2$ \cite{Volmer:2000ek}.  In 2006 and 2007, this domain was revisited \cite{Tadevosyan:2007yd} and a new result was reported, this time at $Q^2=2.45\,$GeV$^2$ \cite{Horn:2006tm}.  However, there was disappointment and surprise, with the collaboration stating that experiment is ``still far from the transition to the $Q^2$ region where the pion looks like a simple quark-antiquark pair.''  This data is represented in Fig.\,\ref{figWPFpi} by the filled circles and squares, drawn from Ref.\,\cite{Huber:2008id}, and the collaboration were comparing with Curve-D in the figure, which is the prediction obtained when $\varphi^{\rm asy}_\pi(x)$, the asymptotic PDA appropriate to the conformal limit of QCD, is used in Eq.\,\eqref{pionUV}.

The confusion was compounded by the fact that JLab's data confirmed the behaviour predicted by a DSE-RL prediction for the pion form factor, computed in 2000 \cite{Maris:2000sk}.  That prediction could not resolve the difficulty because it relied on brute numerical methods and hence could not produce a result at $Q^2 > 4\,$GeV$^2$.  This limit is marked by the vertical dashed-line in Fig.\,\ref{figWPFpi}.  In appearance, however, the shape of the prediction suggested to many that one might never see parton model scaling and QCD scaling violations at the momentum transfers accessible to JLab, even after its upgrade.

This conundrum was recently resolved \cite{Chang:2013nia}.  Using a refinement of known methods \cite{Nakanishi:1963zz,Nakanishi:1969ph}, also employed in the successful analysis of $\varphi_\pi(x;\zeta)$, described in Sec.\,\ref{sec-1}, a reliable prediction of $F_\pi(Q^2)$ is now available on the entire domain of spacelike $Q^2$.  This is Curve-A in Fig.\,\ref{figWPFpi}.  Moreover, the analysis enables correlation of that result with Eq.\,\eqref{pionUV}, using the modern PDA computed in precisely the same framework, which is Curve-C in the figure.  This leading-order, leading-twist QCD prediction, obtained with a pion valence-quark PDA evaluated at a scale appropriate to the experiment, underestimates the full DSE-RL computation by merely an approximately uniform 15\% on the domain depicted.  The small mismatch is explained by a combination of higher-order, higher-twist corrections to Eq.\,\eqref{pionUV} in pQCD on the one hand and, on the other hand, shortcomings in the rainbow-ladder truncation (see Sec.\,\ref{secAbInitio}), which predicts the correct power-law behaviour for the form factor but not precisely the right anomalous dimension (exponent on the logarithm) in the strong coupling calculation.  Hence, disappointment is now transformed into optimism because the comparison of Curves-A and C in Fig.\,\ref{figWPFpi} predicts that the upgraded JLab facility will reveal a maximum at $Q^2 \approx 6\,$GeV$^2$ and an experiment at $Q^2=9\,$GeV$^2$ will see a clear sign of parton model scaling for the first time in a hadron elastic form factor.

The implications of this marked shift in perspective are manifold.  In the foreseeable future one may reasonably expect empirical confirmation of the theory of factorisation in hard exclusive processes, and expose the dominance of hard contributions to the pion form factor for $Q^2>8\,$GeV$^2$.  One will simultaneously find that the normalisation of $F_\pi(Q^2)$ is fixed by a pion wave-function whose dilation with respect to $\varphi_\pi^{\rm asy}(x)$ is a definitive signature of DCSB.  The experiments will thus provide a direct measurement of the strength of DCSB in the Standard Model, the origin of the vast bulk of visible mass.  These will be important pages in a book on the Standard Model, in which the first lines were written forty years ago.

\section{Implications for protons, neutrons and their excited states}
\label{secBaryon}
The understanding at which we have now arrived paves the way for a dramatic reassessment of pictures of proton and neutron structure, which is already well underway \cite{Segovia:2014aza}.  One particular example shall here serve to illustrate that progress.

\begin{figure}[t]
\begin{minipage}[t]{\textwidth}
\begin{minipage}{0.47\textwidth}
\centerline{\includegraphics[clip,width=0.95\textwidth]{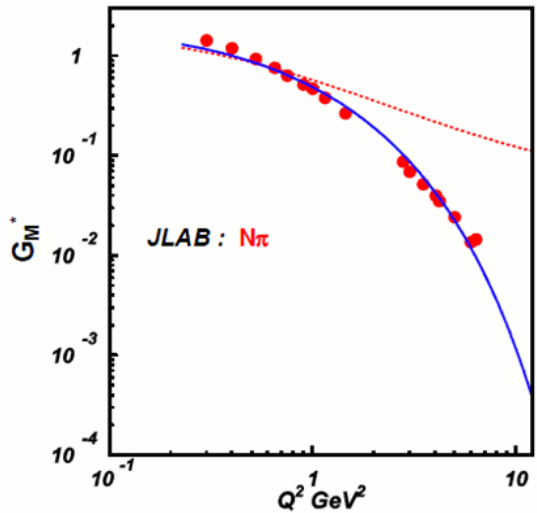}}
\end{minipage}
\begin{minipage}{0.5\textwidth}{\small
\caption{\label{figNDelta} \small
Comparison between CLAS data \cite{Aznauryan:2009mx} on the magnetic $\gamma+N\to\Delta$ transition form factor and a recent theoretical prediction \cite{Segovia:2014aza}.  The dashed curve shows the result that would be obtained if the interaction between quarks in QCD were momentum-independent \cite{Wilson:2011aa}.  The solid curve is obtained with precisely the same QCD-based formulation as was employed in a successful analysis of nucleon elastic form factors, which explains \cite{Cloet:2013gva} the behaviour of the ratio $\mu_p G_E^p(Q^2)/G_M^p(Q^2)$ described in the Introduction.  The experiment-theory comparison establishes that experiments are sensitive to the momentum dependence of the running couplings and masses in QCD; and the theoretical unification of $N$ and $\Delta$ properties highlights the material progress that has been made in constraining the long-range behaviour of these fundamental quantities.  (Figure courtesy of V.\,Mokeev.)}}
\end{minipage}
\end{minipage}
\end{figure}

Given the challenges posed by non-perturbative QCD, it is insufficient to study hadron ground-states alone.  Many novel perspectives and additional insights are provided by nucleon-to-resonance transition form factors, whose behaviour at large momentum transfers can reveal much about the long-range behaviour of the interactions between quarks and gluons \cite{Aznauryan:2012baS}.   Indeed, the properties of nucleon resonances are more sensitive to long-range effects in QCD than are those of hadron ground states.  The lightest baryon resonances are the $\Delta(1232)$-states; and despite possessing a width of 120\,MeV, these states are well isolated from other nucleon excitations.  Hence the $\gamma+N\to\Delta$ transition form factors have long been used to probe strong interaction dynamics.  They excite keen interest because of their use in probing, \emph{inter} \emph{alia}, the relevance of perturbative QCD to processes involving moderate momentum transfers \cite{Carlson:1985mm,Pascalutsa:2006up,Aznauryan:2011qj}; shape deformation of hadrons \cite{Alexandrou:2012da}; and, of course, the role that resonance electroproduction experiments can play in exposing non-perturbative features of QCD \cite{Aznauryan:2012baS}.  Using the ``CLAS'' detector at JLab, precise data on the dominant $\gamma+N\to\Delta$ magnetic transition now reaches to $Q^2 = 8\,$GeV$^2$; an eventuality that poses both great opportunities and challenges for QCD theory, some of which have recently been met, as illustrated in Fig.\,\ref{figNDelta}.

\section{Insights from continuum-QCD}
Perhaps the most important understanding to draw from this contribution is highlighted in connection with Eq.\,\eqref{gtrE}; namely, dynamical chiral symmetry breaking (DCSB) in QCD -- the generation of mass from nothing -- is expressed fundamentally in a near-identity between the dressed-quark one-body problem and the pseudoscalar two-body problem.  As illustrated in Secs.\,\ref{secVacuum}-\ref{secEM}, this has enabled theory to arrive at a detailed understanding of Nature's lightest hadron, \emph{viz}.\ the pion.  The dressed-quark mass function (Fig.\,\ref{gluoncloud}) is known and there is no material model dependence.  Hence, continuum-QCD theory is on the threshold of drawing a complete picture of the pion's light-front valence-quark wave-function, which provides the closest thing in quantum field theory to a quantum mechanical picture of the pion.

In parallel, continuum-QCD is providing predictions of the light-front structure of other light-quark hadrons \cite{Shi:2014uwa,Gao:2014bca} and embarking on analyses of generalised parton distributions for these systems \cite{Chang:2014lva,Mezrag:2014jka}, with a view to providing a unified QCD-based and quantum mechanical picture of the light-quark meson sector of the Standard Model.  Sec.\,\ref{secBaryon} indicates that developments in the baryon sector are equally promising; and Sec.\,\ref{secAbInitio} suggests that continuum-QCD is on the verge of providing genuinely \emph{ab initio} predictions for hadron observables.

\section*{Acknowledgments}
I am grateful to the organisers of the ``37th Brazilian Workshop on Nuclear Physics'' and my hosts in Maresias for their support and gracious hospitality.
I thank, too,
D.~Binosi,
S.\,J.~Brodsky,
L.~Chang,
C.~Chen,
I.\,C.~Clo\"et,
M.-H.~Ding,
B.~El-Bennich,
F.~Gao,
G.~Krein,
T.-S.\,H.~Lee,
Y.-X.~Liu,
C.~Mezrag,
V.~Mokeev,
H.~Moutarde,
J.~Papavassiliou,
M.~Pitschmann,
S.-X.~Qin,
J.~Rodr\'iguez-Quintero,
T.~Sato,
J.~Segovia,
S.\,M.~Schmidt,
R.~Shrock,
C.~Shi,
P.\,C.~Tandy,
S.-L.~Wan,
S.-S.~Xu
and H.-S.~Zong
for valuable discussions during the preparation of this contribution.
This work was supported by U.S.\ Department of Energy, Office of Science, Office of Nuclear Physics, contract no.~DE-AC02-06CH11357.

\section*{References}

\providecommand{\newblock}{}

\end{document}